\documentclass[a4paper, pra, twocolumn,superscriptaddress,showpacs]{revtex4}

\usepackage{amssymb}
\usepackage{amsmath}
\usepackage{epsfig}
\usepackage{color}
\usepackage{graphics, graphicx}
\usepackage{bbold}
\usepackage{psfrag}
\usepackage{mathcomp}
\usepackage{subfigure}
\usepackage{verbatim}
\usepackage{color}
\usepackage[colorlinks,citecolor=blue]{hyperref}

\begin{document}

\date{\today}
\title{Unconventional Fulde-Ferrell-Larkin-Ovchinnikov pairing states in a Fermi gas with spin-orbit coupling}
\author{Fan Wu}
\affiliation{Key Laboratory of Quantum Information, University of Science and Technology of China,
CAS, Hefei, Anhui, 230026, People's Republic of China}
\author{Guang-Can Guo}
\affiliation{Key Laboratory of Quantum Information, University of Science and Technology of China,
CAS, Hefei, Anhui, 230026, People's Republic of China}
\author{Wei Zhang}
\email{wzhangl@ruc.edu.cn}
\affiliation{Department of Physics, Renmin University of China, Beijing 100872, People's Republic of China}
\author{Wei Yi}
\email{wyiz@ustc.edu.cn}
\affiliation{Key Laboratory of Quantum Information, University of Science and Technology of China,
CAS, Hefei, Anhui, 230026, People's Republic of China}

\begin{abstract}
We study the phase diagram in a two-dimensional Fermi gas with the synthetic spin-orbit coupling that
has recently been realized experimentally. In particular, we characterize in detail the properties and the
stability region of the unconventional Fulde-Ferrell-Larkin-Ovchinnikov (FFLO) states in such a system,
which are induced by spin-orbit coupling and Fermi surface asymmetry. We identify several distinct nodal
FFLO states by studying the topology of their respective gapless contours in momentum space. We then
examine the phase structure and the number density distributions in a typical harmonic trapping potential
under the local density approximation. Our studies provide detailed information on the FFLO pairing states
with spin-orbit coupling and Fermi surface asymmetry, and will facilitate experimental detection of these
interesting pairing states in the future.
\end{abstract}

\maketitle

\section{Introduction}
\label{sec:intro}

The recent experimental realization of synthetic spin-orbit coupling (SOC) in ultracold gases of neutral atoms
has stimulated much interest in these systems \cite{gauge2exp,fermisocexp1,fermisocexp2}. As SOC breaks
the inversion symmetry and qualitatively changes the dispersion spectra, a spin-orbit coupled Fermi gas can
exhibit many intriguing properties. For example, in a non-interacting Fermi gas with SOC, the system may
undergo Lifshitz transition as the Fermi surface is tuned \cite{fermisocexp1,soc3}. On the other hand, in an
attractively interacting Fermi gas with SOC, either the topological superfluid state or the exotic nodal superfluid
states related to similar phases in non-centrosymmetric superconducting materials can be stabilized, depending
on the specific type of SOC and the dimensionality of the system \cite{zhang,sato,chuanwei,soc4,soc6,iskin,thermo,
2d2,2d1,melo,wmliu,helianyi,wy2d,xiaosen,wypolaron,carlosnistsoc,iskinnistsoc,puhantwobody,shenoy,wyfflo,melosupp1,melosupp2,puhan3dsoc,XF}.

So far, only an equal Rashba and Dresselhaus (ERD) SOC has been realized experimentally in a three-dimensional
Fermi gas \cite{fermisocexp1,fermisocexp2,xiongjun}, in which case there can be no topological superfluid states
supporting Majorana zero modes in the excitations \cite{carlosnistsoc}. However, it has been pointed out recently
that in the presence of an effective transverse Zeeman field, which can be implemented by tuning the parameters
of the Raman lasers generating the synthetic SOC, exotic pairing states with finite center-of-mass (CoM) momentum
are always more stable than a Bardeen-Cooper-Schrieffer (BCS) pairing state with zero CoM
momentum \cite{puhantwobody,shenoy,wyfflo,hufflo}. The presence of these finite CoM momentum pairing states,
known as Fulde-Ferrell-Larkin-Ovchinnikov (FFLO) states, is a direct consequence of pairing under SOC and
Fermi surface asymmetry, which, in this case, is induced by the effective transverse Zeeman field.
Apparently, the pairing mechanism of these states is quite general, and is {\it different} from that of the
conventional FFLO state, where the pairing takes place between particles on different Fermi surfaces.
Indeed, similar FFLO states have been reported in many other related systems with various types of SOC and in
different dimensions \cite{puhan3dsoc,XF,chuanweifflo,bdg1,bdg2}.

In this work, we study the exotic FFLO states in a two-dimensional (2D) Fermi gas under ERD SOC and cross Zeeman fields,
thus extending our previous work \cite{wyfflo}. In particular, we identify novel nodal FFLO states with topologically distinct gapless
contours in momentum space. Similar nodal FFLO states in systems with different types of SOC and dimensions have been
reported recently, which further implies the generality of the pairing mechanism. We examine in detail the dispersion spectra,
as well as the stability region of these nodal FFLO states on the phase diagram. While the unique dispersion spectra could be
measured via momentum-resolved radio-frequency spectroscopy, the phase transition between these nodal FFLO states could
be identified by measuring the thermodynamic properties, since the difference in the topology of the gapless contours should
lead to different behaviors of low-energy excitations. We then calculate the phase structure and the number density distribution
in a typical harmonic trapping potential under the local density approximation (LDA). We show that under LDA, the first-order phase
transitions on the phase diagram show up as discontinuities in the density distribution. These discontinuities would become smooth
in a realistic experimental setting, but signatures of the phase boundaries should still be observable experimentally, similar to those
in a polarized Fermi gas. The presence of these first-order phase boundaries on the phase diagram and in a trapping potential
suggests the existence of phase-separated states in a uniform gas with a fixed total particle number. Although SOC mitigates the
competition between polarization and pairing by mixing the spin components, for moderate SOC strengths which roughly correspond
to the experimental parameters, first-order phase transitions may still be observed experimentally.

The paper is organized as follows: in Sec.~\ref{sec:model}, we present our mean-field formalism. In Sec.~\ref{sec:thermo},
we study the properties of the thermodynamic potential and examine the stability of BCS-type pairing state against the FFLO
states under the ERD SOC and effective Zeeman fields. We then map out the typical phase diagrams and characterize
the FFLO states in Sec.~\ref{sec:phase}, and demonstrate the existence of a phase-separated state in a uniform gas with a fixed particle number.
In Sec.~\ref{sec:trap}, we calculate typical phase structure and
number density distribution in an isotropic harmonic trap. Finally, we summarize in Sec.~\ref{sec:conclusion}.


\section{Model}
\label{sec:model}

We focus on the ground state properties of a uniform 2D Fermi gas at zero temperature. A mean-field formalism therefore would provide a qualitatively correct picture, although quantitative characterization of the phase boundaries requires a more involved theory which accounts for the fluctuations. Following Ref. \cite{wyfflo}, the effective mean-field Hamiltonian for a uniform two-component Fermi gas under the ERD SOC and effective Zeeman fields can be written in a matrix form under the hyperfine spin basis $\{a_{\mathbf{k},\uparrow},a_{\mathbf{Q}-\mathbf{k},\uparrow}^{\dag},a_{\mathbf{k},\downarrow}, a_{\mathbf{Q}-\mathbf{k},\downarrow}^{\dag}\}^{T}$:
\begin{eqnarray}
H_{\text{eff}}&=&\sum_{\mathbf{k}}M_{\mathbf{k}}+\sum_{\mathbf{k}}\epsilon_{|\mathbf{Q}-\mathbf{k}|}-\frac{|\Delta_{\mathbf{Q}}|^2}{U}\nonumber\\
&=&\frac{1}{2}\sum_{\mathbf{k}} \left[\begin{array}{cccc}\lambda_{\mathbf{k}}^{+}&0&h&\Delta_{\mathbf{Q}}\\ 0&-\lambda_{\mathbf{Q}-\mathbf{k}}^{+}&-\Delta_{\mathbf{Q}}^{*}&-h\\ h&-\Delta_{\mathbf{Q}}&\lambda_{\mathbf{k}}^{-}&0\\ \Delta_{\mathbf{Q}}^{*}&-h&0&-\lambda_{\mathbf{Q}-\mathbf{k}}^{-}\end{array}\right]\nonumber\\
&& + \sum_{\mathbf{k}}\epsilon_{|\mathbf{Q}-\mathbf{k}|}-\frac{|\Delta_{\mathbf{Q}}|^2}{U}.\label{H0}
\end{eqnarray}
Here, $a_{\mathbf{k},\sigma}$ ($a_{\mathbf{k},\sigma}^{\dag}$) are the annihilation (creation) operators for atoms on different hyperfine spin states $\sigma$ with $\sigma=(\uparrow,\downarrow)$. The diagonal terms $\lambda_{\mathbf{k}}^{\pm}=\xi_{\mathbf{k}}\pm\alpha k_{x}\mp h_{x}$, where $\xi_{\mathbf{k}}=\epsilon_{\mathbf{k}}-\mu,\epsilon_{\mathbf{k}}=\hbar^2 k^2/2m$, $\alpha$ is the strength of the spin-orbit coupling, $\mu$ is the chemical potential, and $m$ is the atomic mass. The effective Zeeman fields $h$ and $h_{x}$ are proportional to the Rabi frequency and two-photon detuning associated with the Raman process in the experiment~\cite{gauge2exp}. Following the convention in Ref. \cite{wyfflo}, we refer to $h$ ($h_x$) as the axial (transverse) field. Throughout the work, we will focus on the Fulde-Ferrell-type pairing state, where the CoM momentum of the pairing state is single-valued \cite{fflo}. As a result, the pairing mean field in Eq. (\ref{H0}) is defined as
\begin{align}
\Delta_{\mathbf{Q}}=\frac{U}{\cal S}\sum_{\mathbf{k}}\left\langle a_{\mathbf{Q}-\mathbf{k},\downarrow}a_{\mathbf{k},\uparrow}\right\rangle,
\end{align}
where ${\cal S}$ is the quantization volume in two dimensions. Finally, the bare interaction rate $U$ should be renormalized following the standard relation \cite{zhangpeng}
\begin{align}
\frac{1}{U}=-\frac{1}{S}\sum_{\mathbf{k}}\frac{1}{E_b + 2\epsilon_{\mathbf{k}}},
\end{align}
where $E_b$ is the binding energy of the two-body bound state in a two dimensions without spin-orbit coupling. The binding energy $E_b$ can be adjusted continuously from zero to large positive values, for example, by tuning the magnetic field from the BCS limit to the Bose-Einstein-Condensation (BEC) side of a Feshbach resonance.

\begin{figure}[tbp]
\includegraphics[width=8cm]{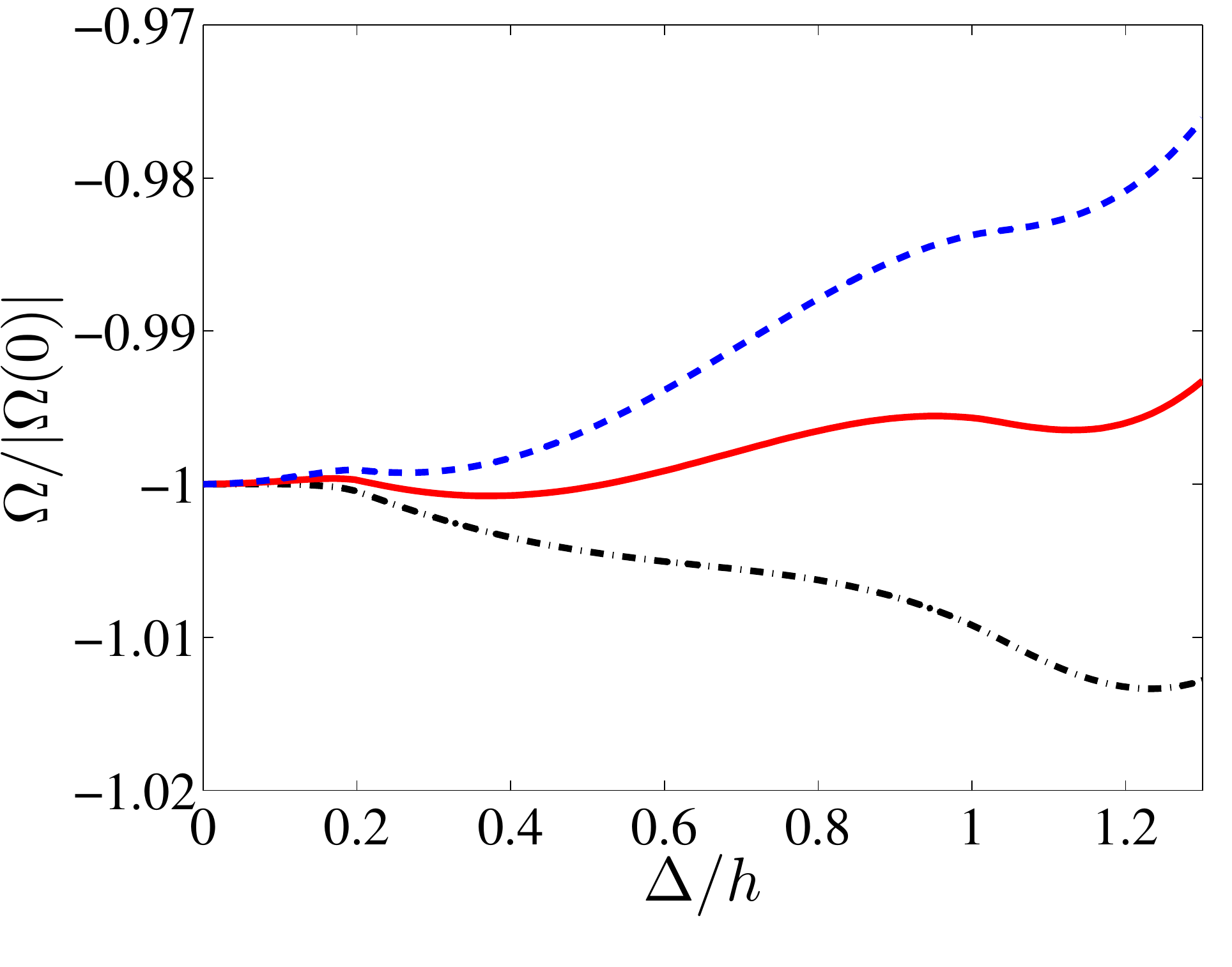}
\caption{(Color online) Typical structure of the thermodynamic potential $\Omega$ as a function of $\Delta$ for $\mathbf{Q}=0$. The parameters for the curves are: $\mu/h=1.374$ (dash-dotted), $\mu/h=1.474$ (solid), $\mu/h=1.624$ (dashed). For all curves, $E_b/h=0.5$, $h_x/h=0.2$, $\alpha k_h/h=0.7$. Here, the effective Rabi-frequency $h$ is taken to be the energy unit, while the unit of momentum $k_h$ is defined as $\hbar^2 k_h^2/(2m)=h$.}\label{TD1}
\end{figure}

It is then straightforward to diagonalize the effective Hamiltonian Eq. (\ref{H0}) and get the thermodynamic potential at zero temperature
\begin{align}
\Omega=\sum_{\mathbf{k}}\xi_{|\mathbf{Q}-\mathbf{k}|}+\sum_{\mathbf{k},\nu} \Theta \left(-E_{\mathbf{k},\nu}^{\eta}\right)E_{\mathbf{k},\nu}^{\eta}-\frac{|\Delta_{\mathbf{Q}}|}{U}.\label{Omg}
\end{align}
Here, the quasi-particle ($\eta = +$) and quasi-hole ($\eta = -$) dispersion $E_{\mathbf{k},\nu}^{\eta}$ ($\nu=1,2$) are the eigenvalues of the matrix $M_{\mathbf{k}}$ in Eq. (\ref{H0}), and $\Theta(x)$ is the Heaviside step function. Without loss of generality, we assume $h$ and $h_x$ are positive, and $\Delta_{\mathbf{Q}}$ is real throughout this work. Before solving for the ground state of the system, in the following, we will first discuss the properties of the thermodynamic potential Eq. (\ref{Omg}), which will turn out to be crucial in determining the correct ground state of the system.

\begin{figure}[tbp]
\includegraphics[width=8cm]{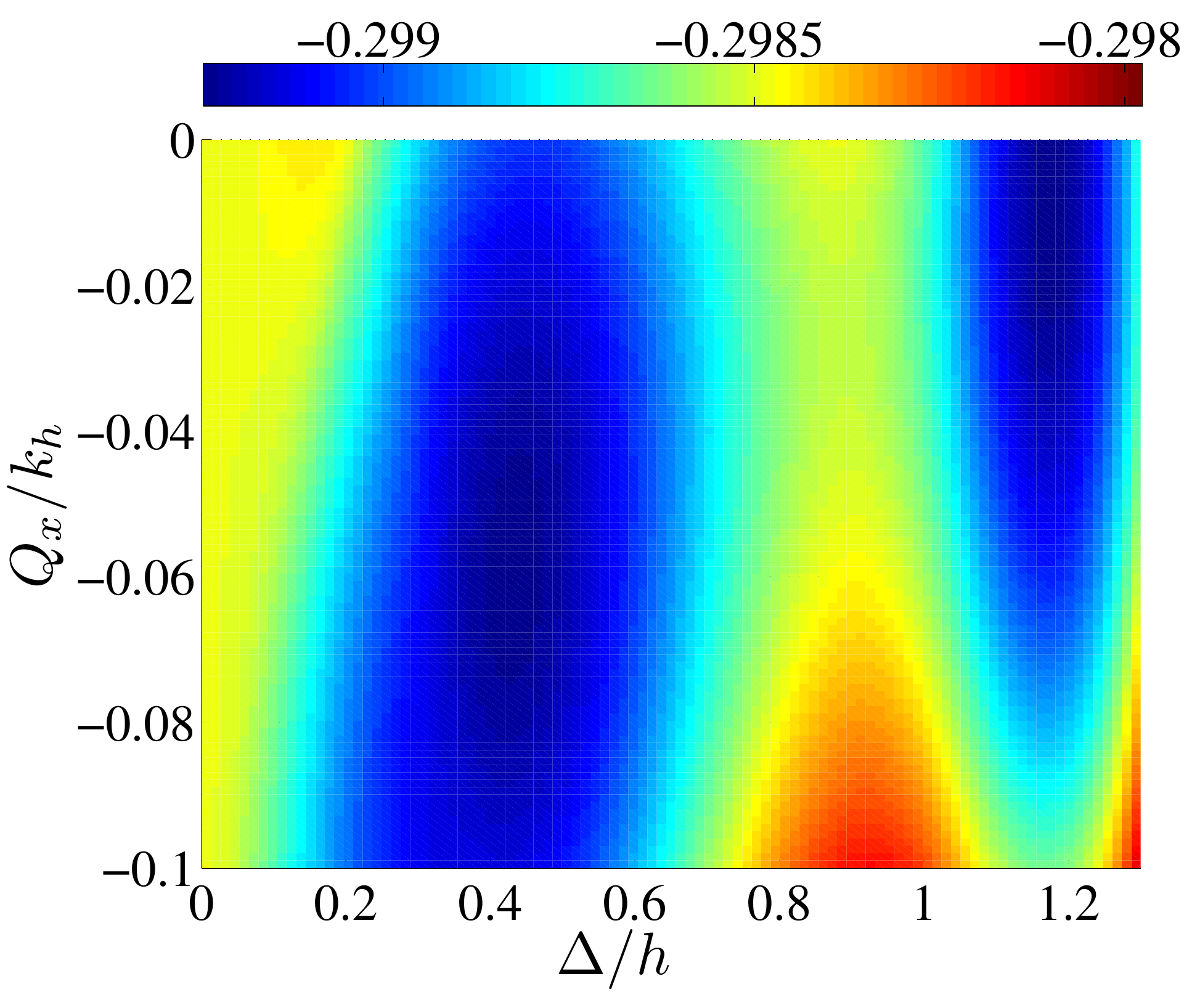}
\caption{(Color online) Contour plot of typical thermodynamic potential landscape in the $\Delta$--$Q_x$ plane for $E_b/h=0.5$, $h_x/h=0.2$, $\alpha k_h/h=0.7$, $\mu/h=1.524$.}\label{TD2}
\end{figure}

\begin{figure}[tbp]
\includegraphics[width=8cm]{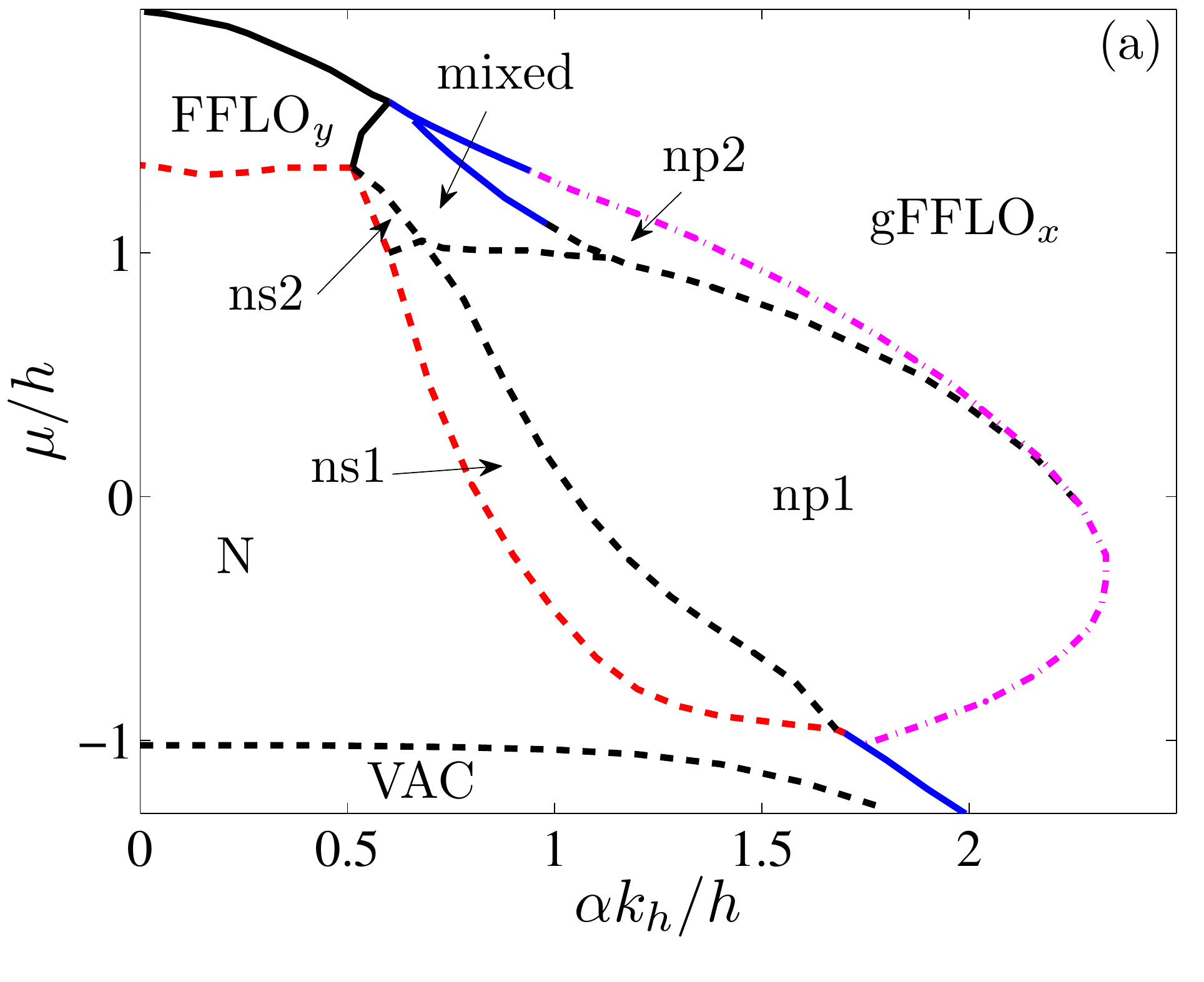}
\includegraphics[width=8cm]{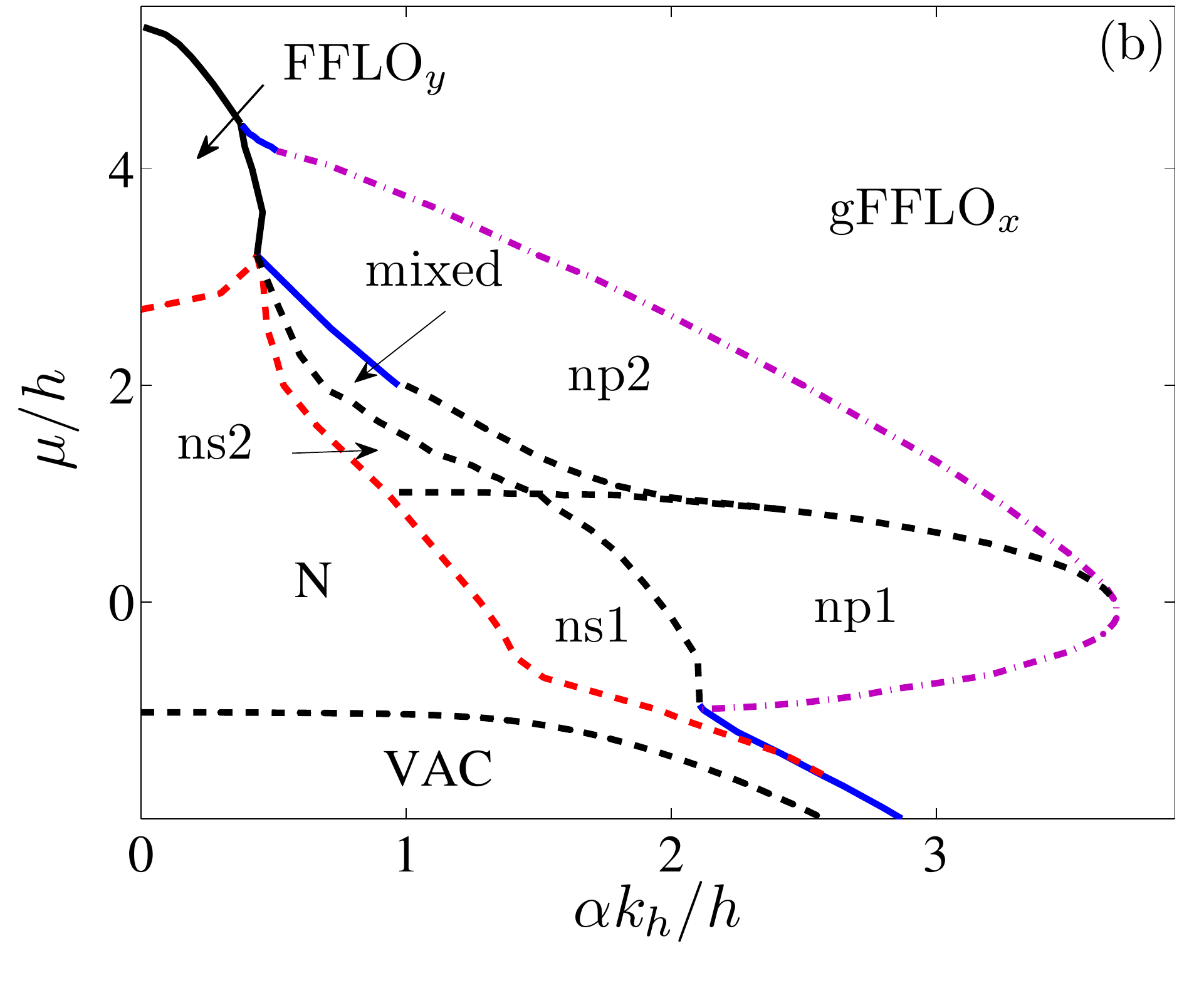}
\caption{(Color online) Typical phase diagrams in the $\alpha$--$\mu$ plane for (a) $E_b/h=0.5$, $h_x/h=0.2$ and (b) $E_b/h=0.2$, $h_x/h=0.2$. Solid curves represent the first order phase boundaries,the dash-dotted curves represent the continuous phase boundary between the fully gapped FFLO state (gFFLO$_x$) and various nodal FFLO$_x$ states, the dashed curves represent continuous phase boundary between different nodal FFLO$_x$ states or the normal state.}\label{aerfa-miu}
\end{figure}


\section{Thermodynamic potential in the presence of ERD SOC and effective Zeeman fields}
\label{sec:thermo}

\begin{figure*}[tbp]
\includegraphics[width=5.5cm]{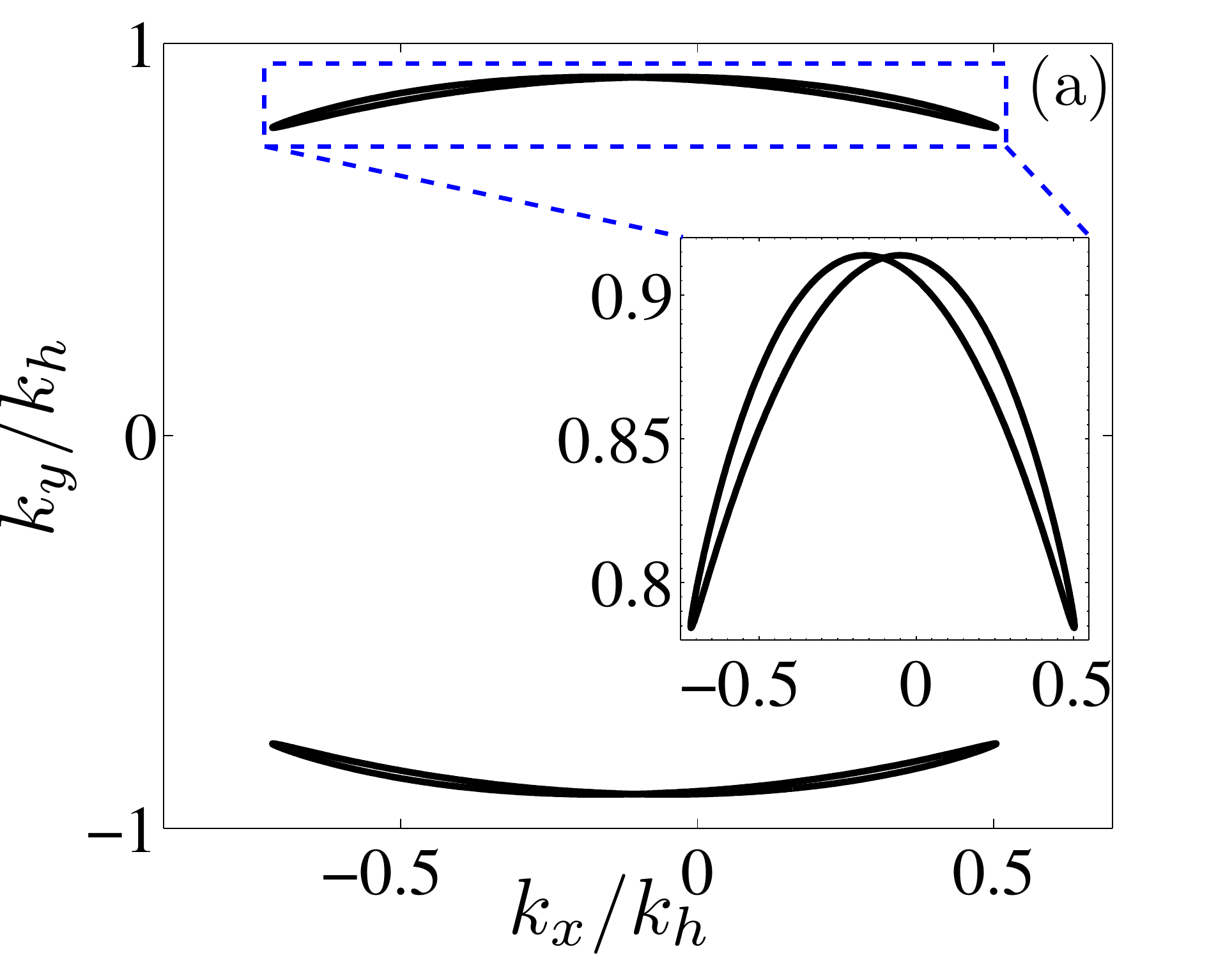}
\includegraphics[width=5.5cm]{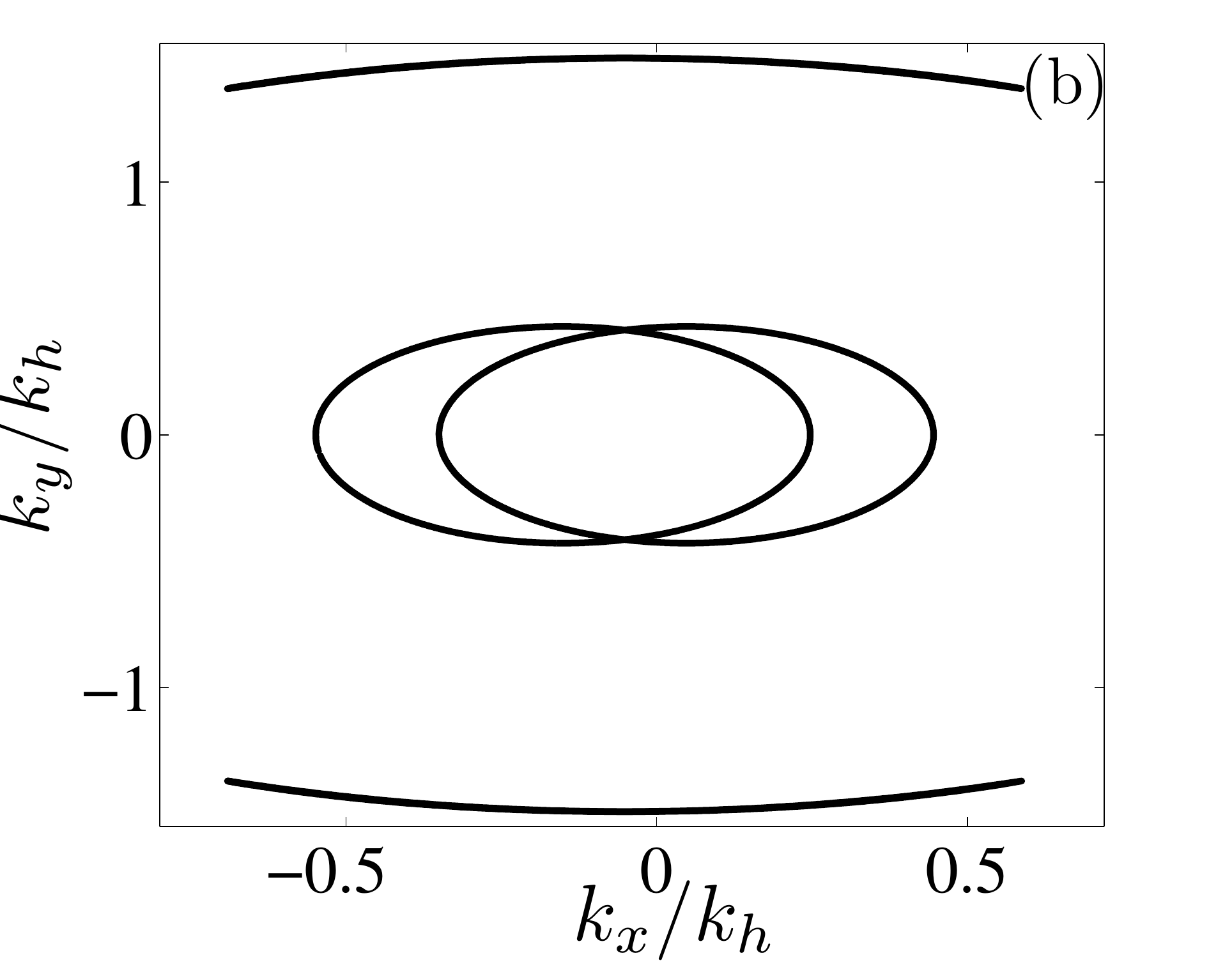}
\includegraphics[width=5.5cm]{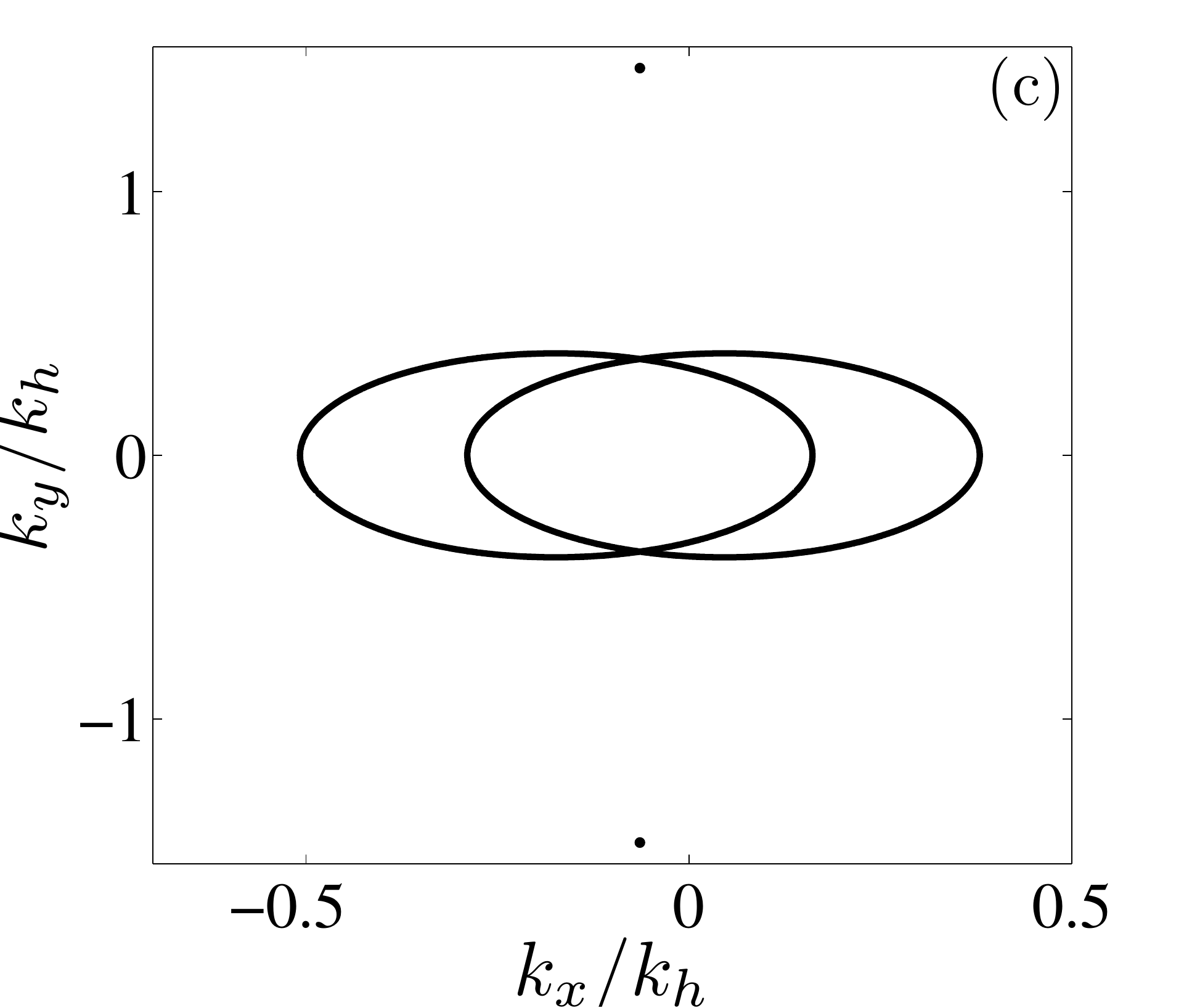}
\includegraphics[width=5.5cm]{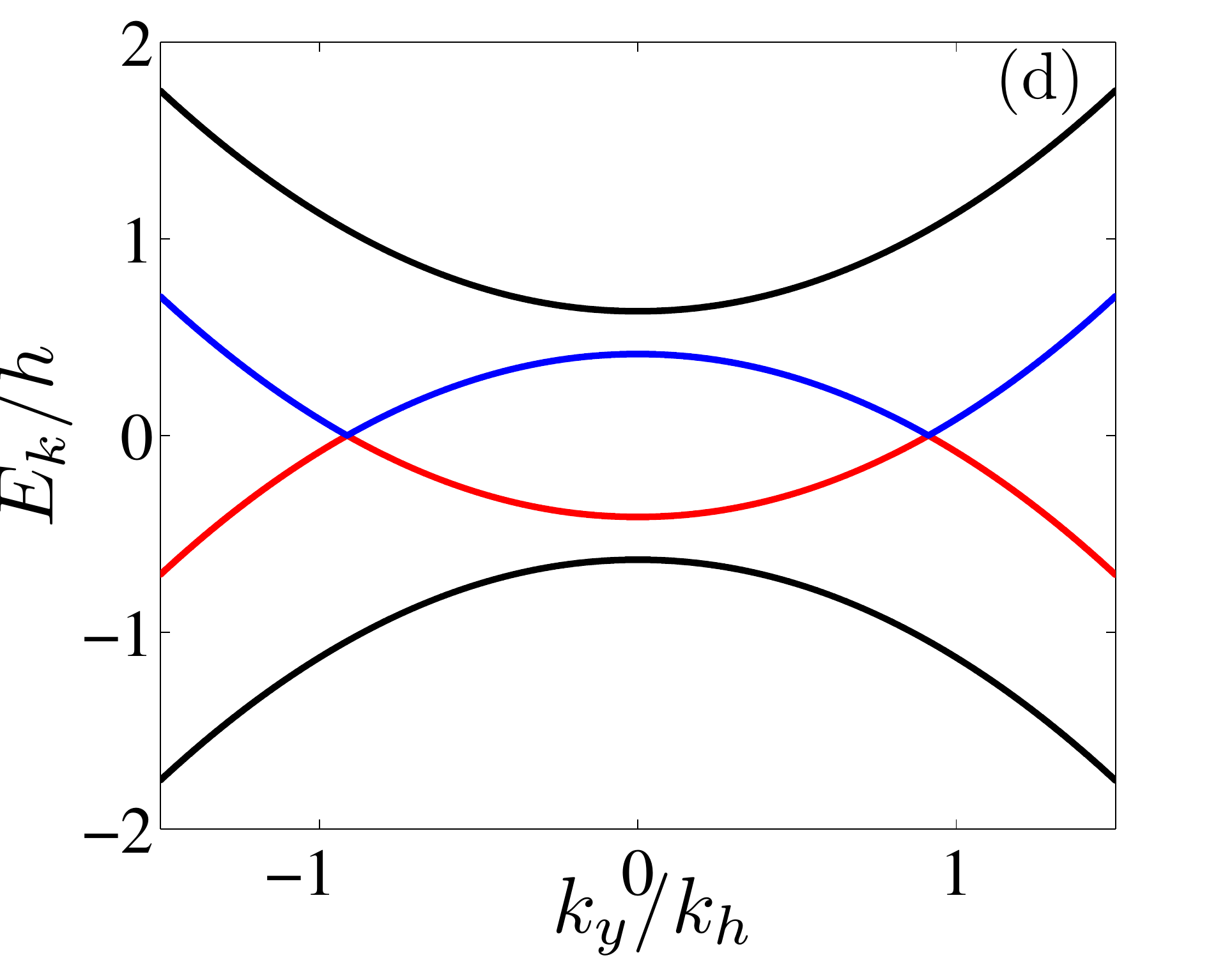}
\includegraphics[width=5.5cm]{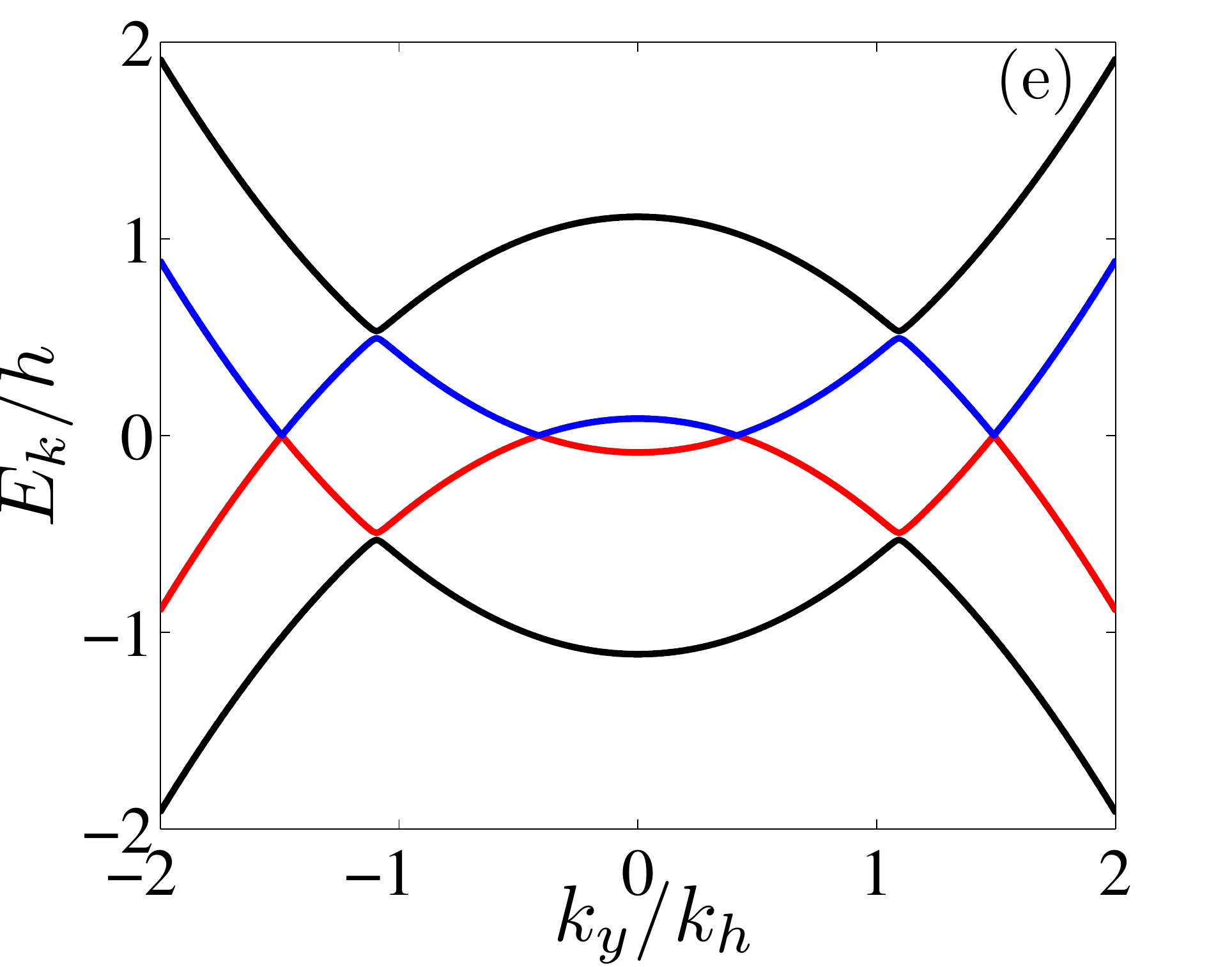}
\includegraphics[width=5.5cm]{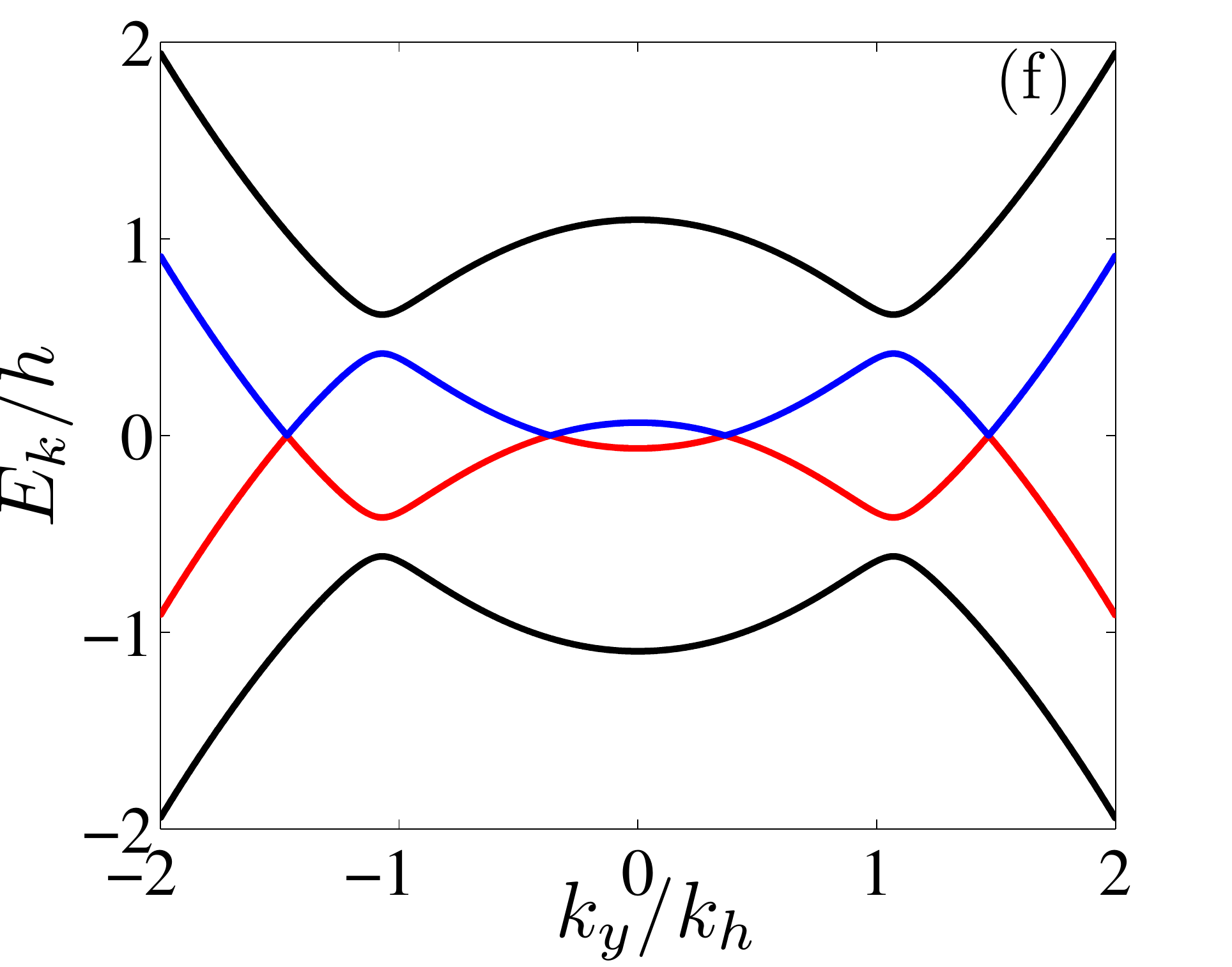}
\caption{(Color online) (a-c) Gapless contours in momentum space for different nodal FFLO$_x$ states with: (a) $\alpha k_h/h=1$, $\mu/h=-0.2$ (ns1); (b) $\alpha k_h/h=0.58$, $\mu/h=1.2$ (ns2); (c) $\alpha k_h/h=0.83$, $\mu/h=1.15$ (mixed). (d-f) Quasi-particle (hole) dispersion spectra along the $k_x=Q_x/2$ axis for the nodal $\text{FFLO}_x$ states that correspond to panels (a-c), respectively.}\label{TP}
\end{figure*}

In this section, we analyze the properties of the thermodynamic potential for a two-dimensional Ferm gas under ERD SOC and effective Zeeman fields. While the zero-temperature thermodynamic potential has been derived in the previous section in terms of the quasi-particle dispersions on the mean-field level, it needs to be evaluated numerically, as, in general, the Hamiltonian (\ref{H0}) cannot be diagonalized analytically. In Fig. \ref{TD1}, we demonstrate typical examples of the thermodynamic potential as a function of the pairing order parameter $\Delta$ for $Q=0$.  Here, an outstanding feature is the possible existence of multiple local minima. As we will show later, similar to the case of a polarized Fermi gas without SOC \cite{preview}, such a structure in the thermodynamic potential landscape suggests that extra care is required when determining the correct ground state of the system, as it is possible to have phase-separated states in a uniform gas with a fixed total particle number.

On the other hand, in the presence of a transverse field $h_x$, it has been reported that the resulting Fermi surface asymmetry, combined with the SOC-induced single-band pairing, should lead to the stabilization of FFLO pairing states with CoM antiparallel to the direction of the effective transverse field. To see this point, we may perform a small $\mathbf{Q}$ expansion of the thermodynamic potential Eq. (\ref{Omg}) around a given local minimum with $Q=0$:
\begin{align}
\Omega\left(\Delta,Q_x\right)=\Omega_0(\Delta)+\Omega_1(\Delta)Q_x+\Omega_2(\Delta)Q_x^2+\mathcal{O}(Q_x^3).
\end{align}
In order to solve for the expansion coefficients $\Omega_i(\Delta)$'s, we first perform an expansion of the eigenvalues of the matrix $M_{\mathbf{k}}$ in the Hamiltonian (\ref{H0})
\begin{align}
E_i=E_{i0}+\delta_{i1}Q_x+\delta_{i2}Q_x^2+\mathcal{O}(Q_x^3).
\end{align}
Then, by employing the identity
\begin{equation}
{\rm tr} \left( M_{\mathbf{k}}^n \right) = \sum_i E_i^n
\end{equation}
and by matching coefficients, we can obtain the following equations for $\delta$'s
\begin{eqnarray}
\left[\begin{array}{cccc}
1 & 1 & 1 & 1\\
E_{10} & E_{20} & E_{30} & E_{40}\\
E_{10}^2 & E_{20}^2 & E_{30}^2 & E_{40}^2\\
E_{10}^3 & E_{20}^3 & E_{30}^3 & E_{40}^3
\end{array}\right]
\left[\begin{array}{c}
\delta_{11}\\
\delta_{21}\\
\delta_{31}\\
\delta_{41}
\end{array}\right] =
\left[\begin{array}{c}
T_{11}\\
T_{21}/2\\
T_{31}/3\\
T_{41}/4
\end{array}\right].\label{ec}
\end{eqnarray}
Here, $T_{ni}$ are the coefficients of the expansion
\begin{align}
{\rm tr}(M_{\mathbf{k}}^n)\approx T_{n0}+T_{n1}Q_x+T_{n2}Q_x^2+\mathcal{O}(Q_x^3).
\end{align}
It is then straightforward to solve Eq. (\ref{ec}) numerically and evaluate $\delta_{i1}$, which are related to $\Omega_1(\Delta)$  via Eq. (\ref{Omg}).

We find that while the first-order expansion coefficient $\Omega_1(\Delta)$ vanishes for $h_x=0$ and finite $\Delta$, it is typically non-zero and has the opposite sign to $h_x$ for non-zero transverse field. Therefore, as $h_x$ is switched on, a pairing state with zero CoM momentum would be shifted onto the finite $\mathbf{Q}$ plane, and the BCS pairing state can no longer be the ground state of the system for finite $h_x$.

As a conclusion of the analysis in this section, we see that the ground state is in general the result of the competition between various FFLO states. This is illustrated in Fig. \ref{TD2}, where we show the contour of the thermodynamical potential in the plane of $\Delta$--$Q_x$. It is also important to keep in mind that we have only analyzed above the possibility of an FFLO state with the CoM momentum lying along the $x$-axis. As we will show later, an FFLO state with CoM momentum in the $y$-axis can also be stabilized.


\section{Nodal FFLO states and their stability region}
\label{sec:phase}

With the understanding of the thermodynamic potential, we can now study the phase diagram of the system. As the first step, we fix the chemical potential $\mu$ and look for the global minimum of the thermodynamic potential in order to get the ground state of the system. With a given chemical potential, except for the first-order phase boundaries, there is only one global minimum, hence the complication of phase separation can be avoided.

We map out the typical phase diagrams in the $\alpha$--$\mu$ plane for $E_b/h=0.5$ [Fig. \ref{aerfa-miu}(a)] and $E_b/h = 0.2$ [Fig. \ref{aerfa-miu}(b)], where the traverse field is chosen as $h_x/h=0.2$. Under the local density approximation (LDA), the chemical potential decreases from the trap center to its edge. Therefore, the typical phase structure in a trapping potential under LDA can be directly identified from our phase diagram. Consistent with our previous analysis, all pairing states on the phase diagram with zero CoM momentum are replaced by FFLO states with finite CoM momentum.

There are three qualitatively different classes of FFLO states on the phase diagram: the FFLO$_y$ state, the fully gapped FFLO state (gFFLO$_x$),  and the various nodal FFLO states (nFFLO$_x$). The FFLO$_y$ state is basically the conventional FFLO state in two dimensions dressed by the ERD SOC. In fact, in the absence of SOC, FFLO states can be stabilized in a 2D polarized Fermi gas over certain parameter regions. There, the rotational symmetry does not single out any particular direction for the pair CoM momentum. In the presence of the ERD SOC and cross Zeeman fields, however, we find numerically that the anisotropy of the system favors a pairing state with CoM momentum perpendicular to the direction of SOC. Indeed, in these parameter regions, we note that the FFLO state with CoM momentum along the direction of SOC is metastable.

On the other hand, both the gFFLO$_x$ and the nodal FFLO$_x$ states can be understood as a shift of the local minima of the thermodynamic potential onto the finite $\mathbf{Q}$ plane induced by the transverse field. When this shift in phase space is not too large, most of the properties of the final FFLO$_x$ state are similar to those of the original pairing state with zero CoM momentum. For example, the phase boundary between the gFFLO$_x$ state and the nodal FFLO$_x$ states (the dash-dotted curve in Fig. \ref{aerfa-miu}) is similar to the phase boundary between the fully gapped superfluid state and the nodal superfluid state on the $h_x=0$ phase diagram (see Ref. \cite{wyfflo}). In particular, the nodal FFLO$_x$ states along this continuous phase boundary either have two (np1) or four (np2) separate gapless points in momentum space, which are the counterparts to the two different nodal superfluid states in the absence of the transverse field. When the shift in phase space becomes larger, novel nodal FFLO$_x$ states with topologically different nodal structure in momentum space appear, resulting in a rich phase structure in the middle of the phase diagrams in Fig. \ref{aerfa-miu}.

To characterize these new nodal FFLO$_x$ states, we calculate their corresponding gapless contours in momentum space [see Fig. \ref{TP}(a-c)]. Typically, we can have nodal FFLO$_x$ states with two (ns1) or four (ns2) separate gapless contours in momentum space. In addition, between the np2 and ns2 states, we also have a mixed region, where two disconnected gapless contours co-exist with two separate gapless points. We have also shown in Fig. \ref{TP}(d-f) the quasi-particle (hole) dispersion spectra of the ns1, ns2 and the mixed states, respectively, which are consistent with the corresponding gapless contours in Fig. \ref{TP}(a-c). In principle, one may probe these dispersion spectra experimentally using momentum resolved radio-frequency spectroscopy, from which the topology of the momentum space gapless contours of the various nodal FFLO$_x$ states can be probed. Finally, we note that as the system approaches the BCS limit, i.e., as $E_b$ decreases, the stability region of the nodal FFLO states, FFLO$_y$ included, increases dramatically (see Fig. \ref{aerfa-miu}).

\begin{figure}[tbp]
\includegraphics[width=8cm]{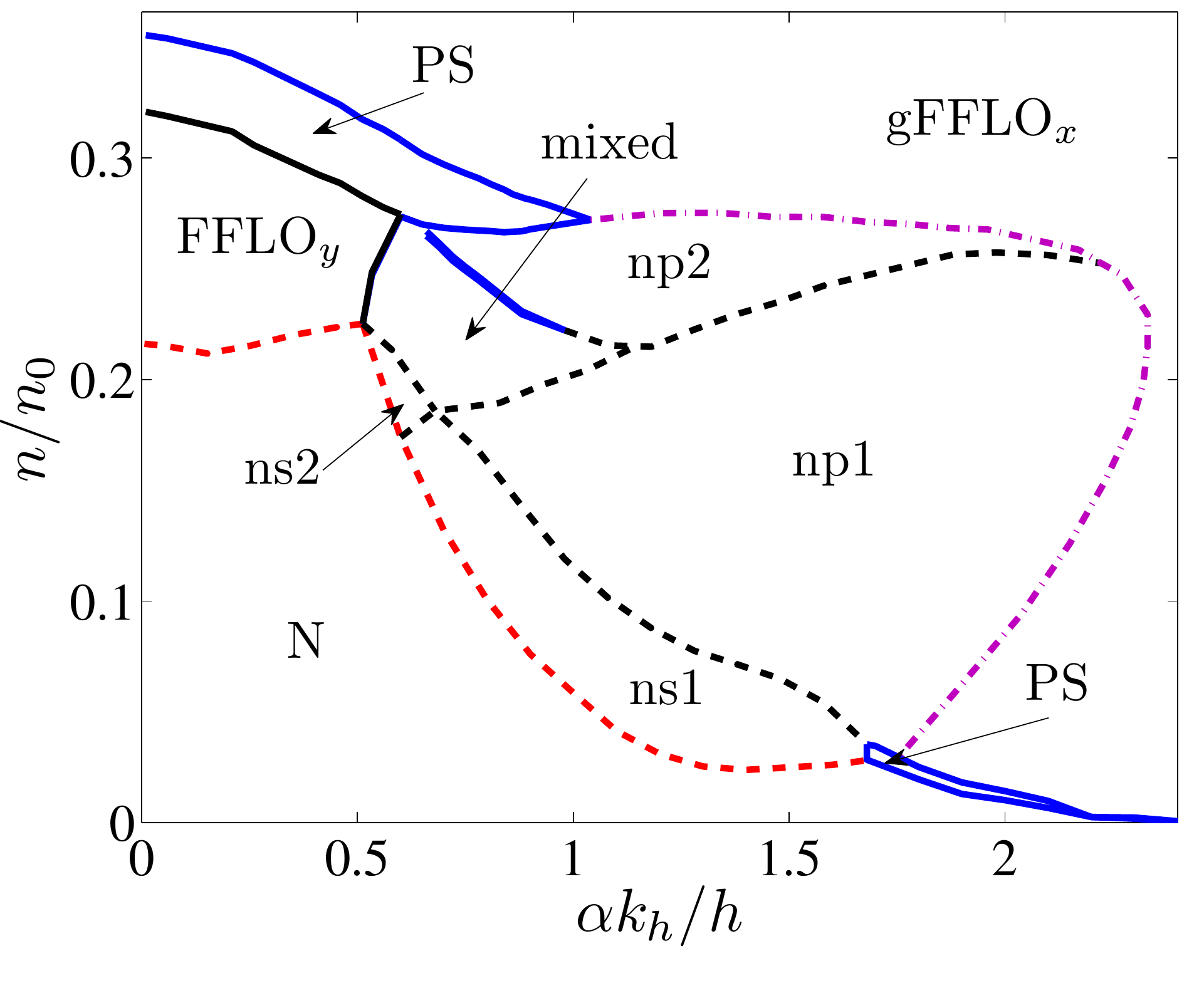}
\caption{(Color online) Typical phase diagram in the $\alpha$--$n$ plane for $E_b/h=0.5$ and $h_x/h=0.2$. Solid curves represent first order phase boundaries, dash-dotted curves represent continuous phase boundary between the fully gapped FFLO state (gFFLO$_x$) and various nodal FFLO$_x$ states, and dashed curves represent continuous phase boundaries between different nodal FFLO$_x$ states or the normal state. We take the unit of number density as $n_0 = k_h^2$.}\label{aerfa-n}
\end{figure}

With the understanding of the $\alpha$--$\mu$ phase diagrams, we then map out the phase diagram on the
$n$--$\alpha$ plane by evaluating the particle number densities at the phase boundaries.
As we have discussed in Sec.~\ref{sec:thermo}, the first-order phase boundaries originate from the
existence of multiple local minima in the thermodynamic potential landscape. Thus, by tuning through
these boundaries, the number density of the ground state exhibits a discontinuous variation as jumping
from one local minimum to another. As a consequence, for a uniform system with a fixed total particle density,
one must explicitly take the phase-separated states into account to get the correct ground state of the system.
Indeed, phase-separated (PS) regions can be easily identified on the $\alpha$--$n$ plane, as shown in \
Fig. \ref{aerfa-n}. Within these regions, the number equations do not support solutions that correspond to the global minimum of the thermodynamic potential. The ground state of the system in this case is a mixture of the
states on both sides of the PS region.


\section{Phase structure in a trapping potential}
\label{sec:trap}

\begin{figure}[tbp]
\includegraphics[width=8cm]{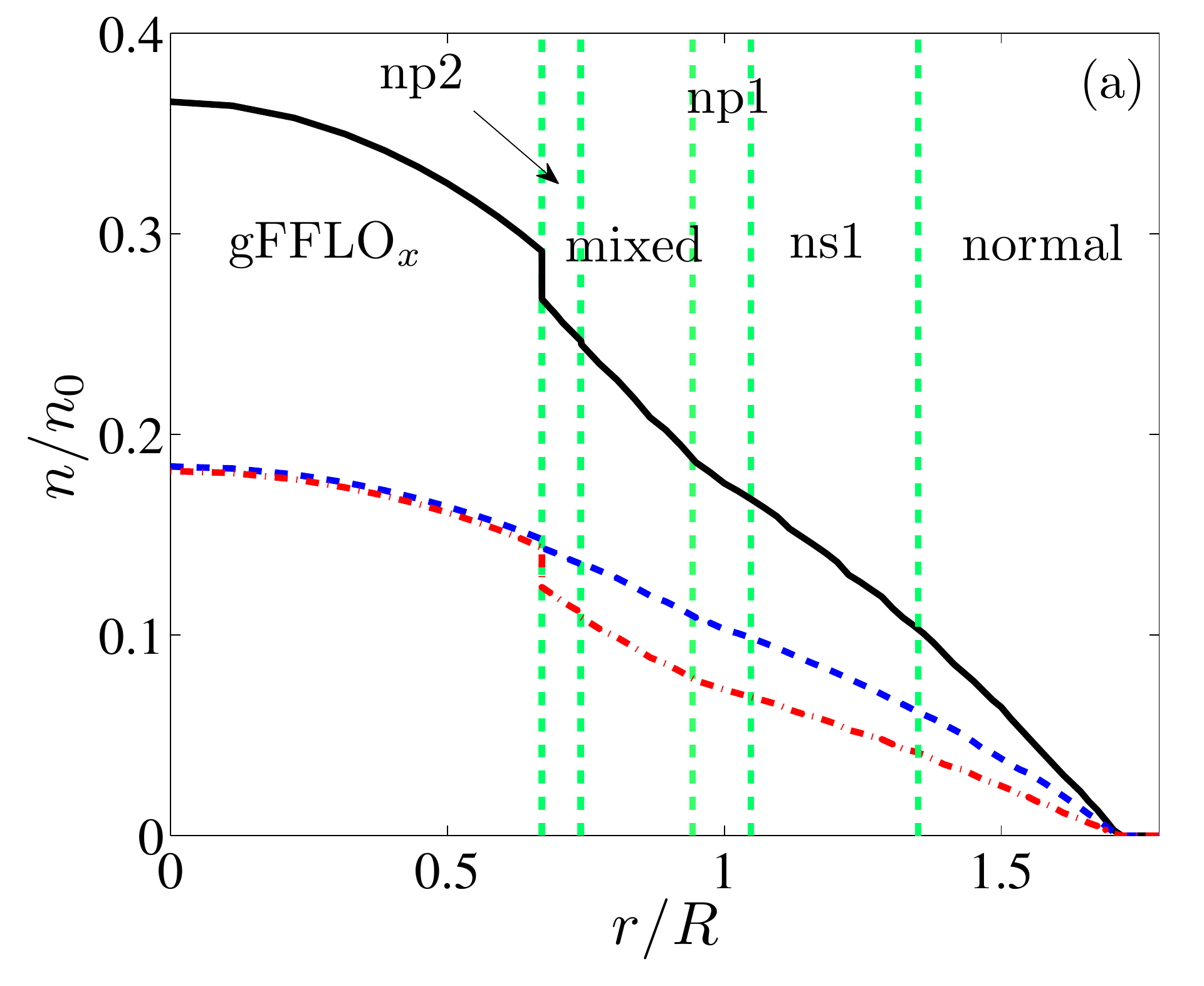}\label{tp0}
\includegraphics[width=8cm]{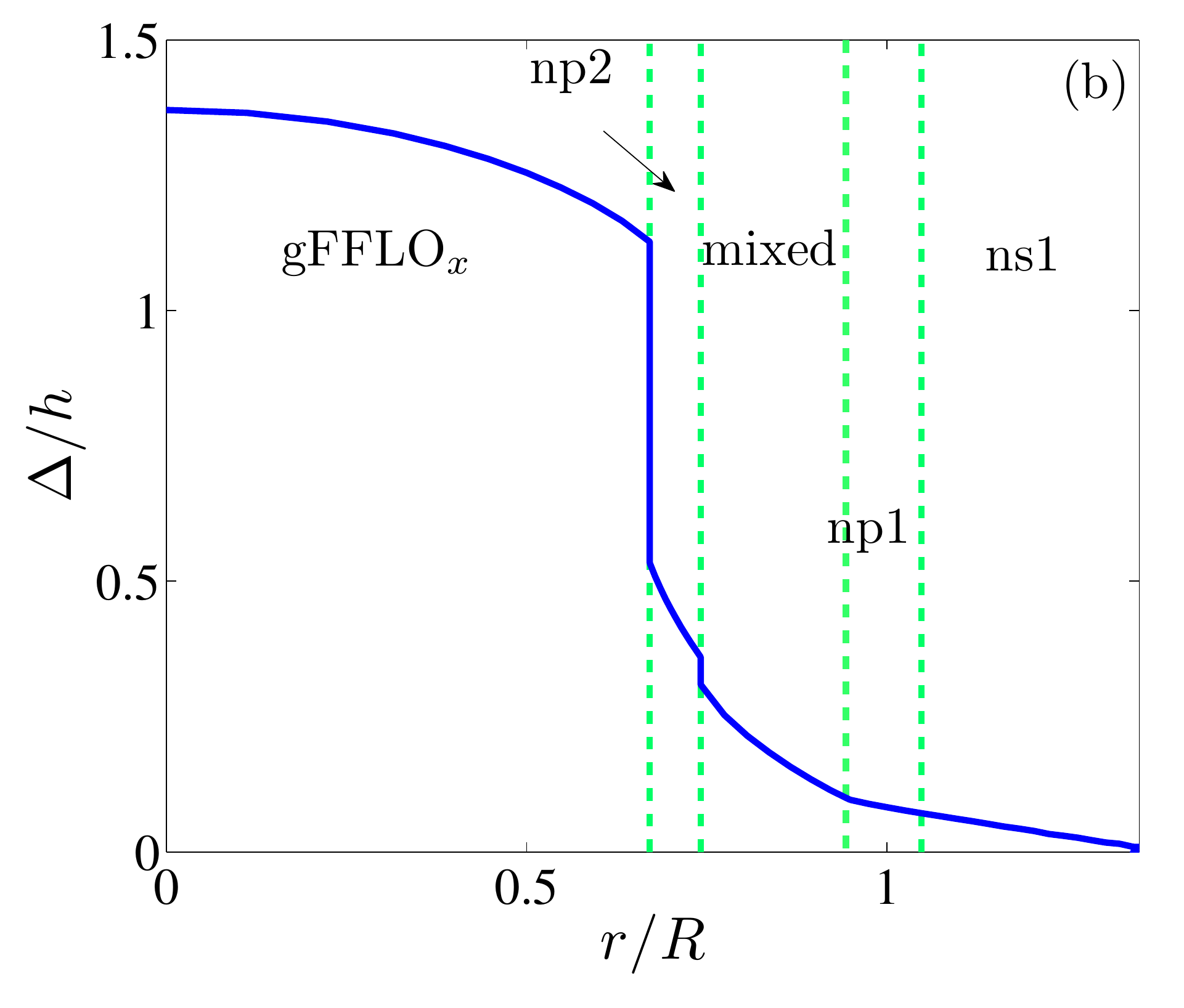}\label{tpD}
\includegraphics[width=8cm]{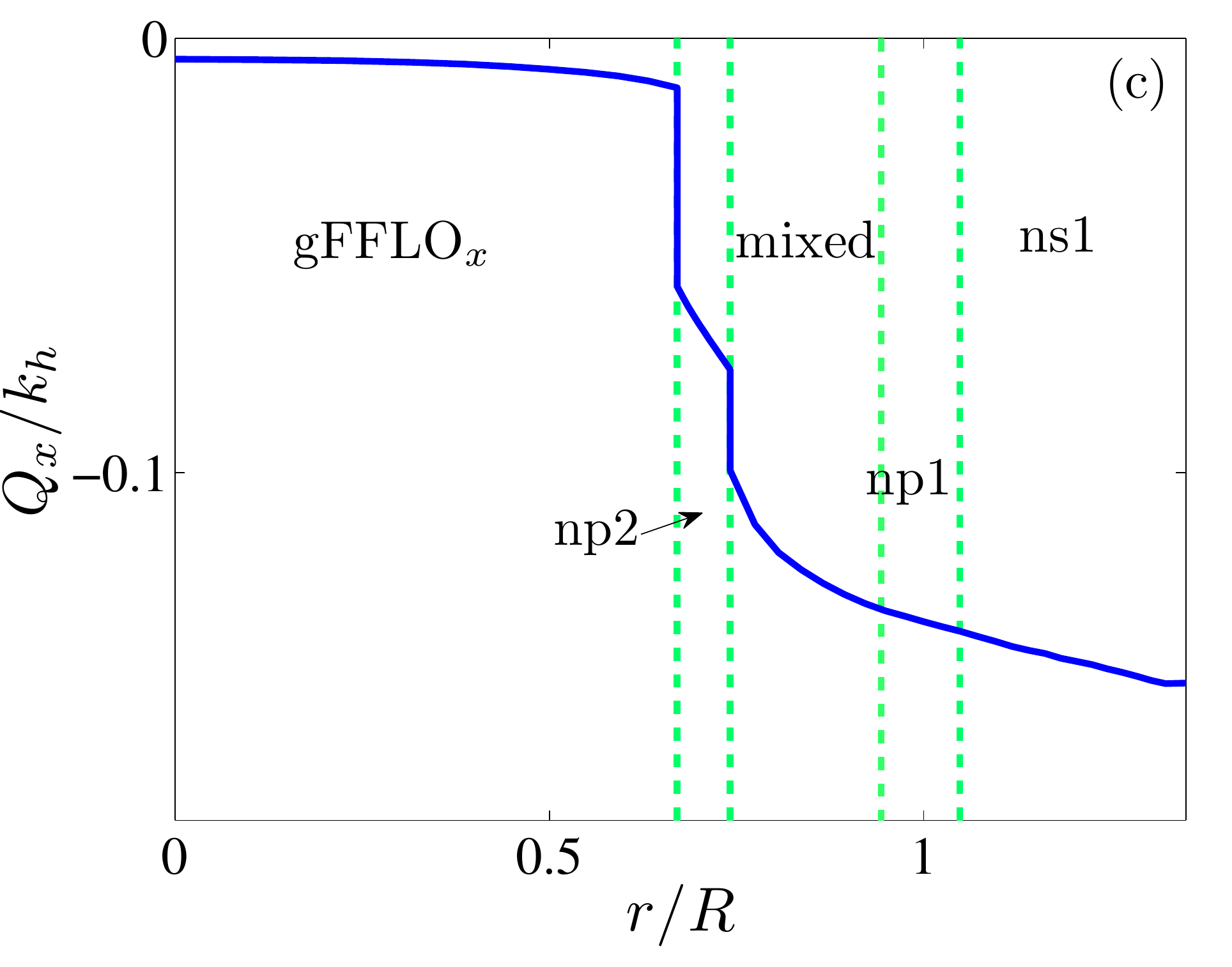}\label{tpQ}
\caption{(Color online)(a) Number density distribution in a typical trapping potential. The solid curve is the total number density distribution, the dashed (dash-dotted) curve represents the density profile of the spin-up (spin-down) component. The units for the number density $n_0$ and for the distance from the trap center $R$ are defined in the text. (b) Variation of the pairing order parameter $\Delta_{Q}$ as a function of the distance from the trap center. (c) Variation of the CoM momentum $Q_x$ of the pairing state as a function of the distance from the trap center. }\label{TRAP}
\end{figure}

With the knowledge of the phase diagram for a homogeneous system, it is now straightforward to get the typical phase structure in a trap under the LDA.  As an example, we consider a 2D Fermi gas in an isotropic harmonic trapping potential under ERD SOC and cross Zeeman fields. To make connection with the experiments, we take typical experimental parameters \cite{gauge2exp,fermisocexp1,fermisocexp2} with trapping frequency $\omega\sim4\pi\times50$Hz and $h\sim2\pi\hbar\times8.3$kHZ. With the energy unit $h$, the harmonic trapping potential in the dimensionless form can be written as:
\begin{align}
\frac{V(\mathbf{r})}{h}=\frac{r^2}{R^2},
\end{align}
where the length unit is defined as $R=\sqrt{{2h}/{m\omega^2}}$. The dimensionless number equations can be written as
\begin{align}
N = \int{n(\mathbf{r})d^2\mathbf{r}} = k_h^2R^2\int{\tilde{n}(\tilde{\mathbf{r}})d^2\tilde{\mathbf{r}}},
\end{align}
where $\tilde{n}=n/n_0, \tilde{\mathbf{r}}=\mathbf{r}/R$, and the unit of number density $n_0 = k_h^2$.

In Fig. \ref{TRAP}, we show the typical phase structure in a harmonic trapping potential, with a total particle number $N\sim 4\times 10^4$. Consistent with the phase diagram in the $\alpha$--$\mu$ plane, the various phases form a shell structure in the trap. The first-order boundaries in Fig. \ref{aerfa-miu} manifest themselves as abrupt changes in the spatial distribution of various parameters (see Fig. \ref{TRAP}). These discontinuous variations are the artifacts of LDA, and will become smooth in a real experiment. However, similar to a polarized Fermi gas without SOC, the first-order phase boundary should still leave experimentally detectable signatures in the density profile of a trapped gas \cite{ketterlepolarized,huletpolarized}.


\section{Summary}
\label{sec:conclusion}

We have studied the properties of the FFLO state in a 2D Fermi gas under the ERD SOC and effective Zeeman fields at zero temperature. Due to the presence of SOC and Fermi surface asymmetry, the zero CoM momentum pairing states are no longer the ground state of the system. Depending on the parameters, the resulting FFLO states can have different CoM momentum and can either be fully gapped or feature gapless excitations. We show that depending on the topology of the gapless contours in the momentum space, one may define several different types of nodal FFLO states, which can be identified by a direct measurement of the quasi-particle dispersion spectra, or by measuring the thermodynamic properties of the system. We also characterize the stability of the different FFLO states in a typical harmonic trapping potential. In particular, we explicitly show that due to the interplay of SOC and Zeeman fields, a spatial phase separation may occur in a trapped gas. The resulting first-order phase boundaries leave observable signatures in the {\it in-situ} density profiles. Our study is relevant to the ongoing efforts in clarifying the properties of the pairing states in a spin-orbit coupled Fermi gas, and provides details which should facilitate future experimental observation of these states.

\acknowledgments
This work is supported by NFRP (2011CB921200, 2011CBA00200), NKBRP (2013CB922000), NNSF (60921091), NSFC (11105134, 11274009), SRFDP (20113402120022), the Fundamental Research Funds for the Central Universities (WK2470000006), and the Research Funds of Renmin University of China (10XNL016).


\end{document}